\pdfoutput=1

\documentclass[11pt]{article}

\usepackage{EMNLP2023}

\usepackage{times}
\usepackage{latexsym}

\usepackage[T1]{fontenc}

\usepackage[utf8]{inputenc}

\usepackage{microtype}

\usepackage{inconsolata}

\usepackage{graphicx}
\usepackage{multirow}
\usepackage{booktabs}
\usepackage{amssymb}
\usepackage{float}
\usepackage{listings}
\usepackage{xcolor}

\definecolor{codegreen}{rgb}{0,0.6,0}
\definecolor{codegray}{rgb}{0.5,0.5,0.5}
\definecolor{codepurple}{rgb}{0.58,0,0.82}
\definecolor{backcolour}{rgb}{0.9843,0.9882,0.9609}

\lstdefinestyle{mystyle}{
    backgroundcolor=\color{backcolour},   
    commentstyle=\color{codegreen},
    keywordstyle=\color{magenta},
    numberstyle=\tiny\color{codegray},
    stringstyle=\color{codepurple},
    basicstyle=\ttfamily\footnotesize,
    breakatwhitespace=false,         
    breaklines=true,                 
    captionpos=b,                    
    keepspaces=true,                 
    numbers=left,                    
    numbersep=5pt,                  
    showspaces=false,                
    showstringspaces=false,
    showtabs=false,                  
    tabsize=2
}
\title{OpenVNA: A Framework for Analyzing the Behavior of Multimodal Language Understanding System under Noisy Scenarios}

\author{Ziqi Yuan\textsuperscript{\rm 1, 2}, Baozheng Zhang\textsuperscript{\rm 1, 3, 5}, Hua Xu\textsuperscript{\rm 1, 5}\thanks{\quad  Hua Xu is the corresponding author. Email: xuhua@tsinghua.edu.cn},  Zhiyun Liang\textsuperscript{\rm 1, 4}, \and Kai Gao\textsuperscript{\rm 3}\\
         {\small\textsuperscript{\rm 1}  State Key Laboratory of Intelligent Technology and Systems, Department of Computer Science and Technology, Tsinghua University}\\
         {\small\textsuperscript{\rm 2} Beijing National Research Center for Information Science and Technology (BNRist)}\\
         {\small\textsuperscript{\rm 3} School of Information Science and Engineering, Hebei University of Science and Technology}\\
         {\small\textsuperscript{\rm 4} College of Information and Electrical Engineering, China Agricultural University}\\
         {\small\textsuperscript{\rm 5} Samton (Jiangxi) Technology Development Co., Ltd}\\
         {\small \texttt{xuhua@tsinghua.edu.cn}}\\
         }

\begin{document}
\maketitle
\begin{abstract}

We present OpenVNA, an open-source framework designed for analyzing the behavior of multimodal language understanding systems under noisy conditions. OpenVNA serves as an intuitive toolkit tailored for researchers, facilitating convenience batch-level robustness evaluation and on-the-fly instance-level demonstration. It primarily features a benchmark Python library for assessing global model robustness, offering high flexibility and extensibility, thereby enabling customization with user-defined noise types and models. Additionally, a GUI-based interface has been developed to intuitively analyze local model behavior. In this paper, we delineate the design principles and utilization of the created library and GUI-based web platform. Currently, OpenVNA is publicly accessible at \url{https://github.com/thuiar/OpenVNA}, with a demonstration video available at \url{https://youtu.be/0Z9cW7RGct4}.
\end{abstract}

\section{Introduction}

The Multimodal Language Understanding (MLU) task aims to empower artificial intelligence agents with the capability to comprehensively understand human communication, discerning the speaker's affective states \cite{baltruvsaitis2018multimodal, soleymani2017survey} and intentions \cite{zhang2022mintrec}. Despite the proliferation of multimodal large language models has yielded remarkable achievements \cite{li2023videochat, video_chatgpt, zhang2023video}, their application in real-world scenarios is still under development. 

Analyzing the behaviors of MLU systems under carefully constructed noise offers a potential avenue for researchers to gain deeper insights into the possible limitations and underlying mechanisms of current MLU systems \cite{multibench, liang2022foundations}. Specifically, by evaluating the global behavior of MLU system under homologous manually constructed noise, researchers can ensure their model operate effectively in practical usage scenarios. Moreover, analyzing the local behavior of the MLU system under customized perturbed instances provides an understanding of how the model makes decisions, pinpointing which aspects of multimodal signals are pivotal for prediction. In recent years, researchers in this field have faced obstacles in imitating real-world noise in multimodal systems and quantitatively assessing the global robustness of MLU methods \cite{ma2022multimodal, hazarika2022analyzing}. Due to the lack of open-source noise injection toolkits and evaluation benchmarks, researchers frequently provide the model performance under specific simulated noise, leading to inequitable comparisons and susceptibility to overfitting on such noise patterns \cite{yuan2023noise}.

\begin{figure*}
    \centering
    \includegraphics[width=0.9\linewidth]{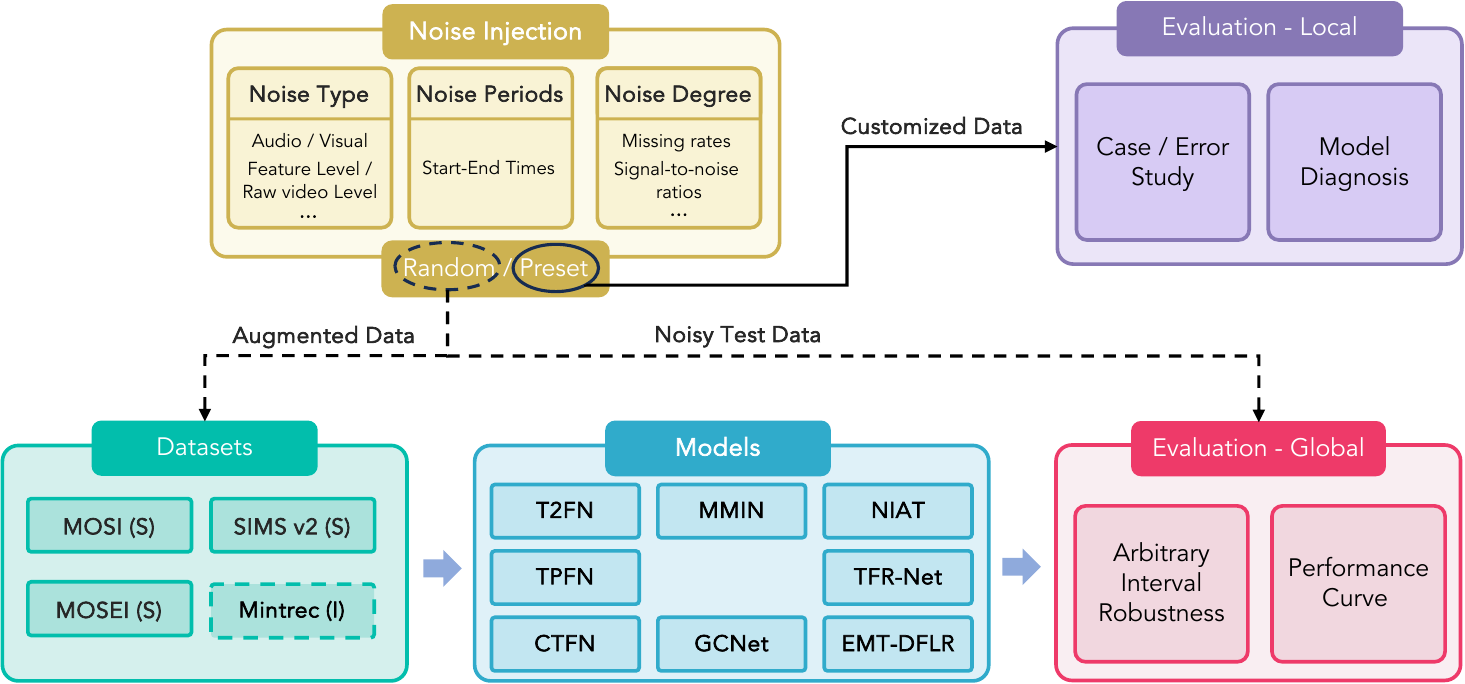}
    \caption{OpenVNA is an comprehensive Python library for analyzing MLU system behavior under noise, which consists of noise injection module, datasets module, models module, instance-level (local) and batch-level (global) evaluation module. For instance level model behavior evaluation, an online platform is also available.}
    \label{fig: structural}
\end{figure*}

To bridge the gap between simulated scenario and real world multimodal noise, benchmark current approaches, and provide intuitive local behavior analysis, we introduce OpenVNA, an open-source framework dedicated to analyze MLU system behavior under noisy scenario. 
The composition of the OpenVNA and the interconnections among its components are illustrated in Figure \ref{fig: structural}.
Firstly, OpenVNA serves as a Python Library providing easy-to-use Application Programming Interfaces (APIs) for noise simulation, model reproduction, and global robustness evaluation. Presently, it incorporates fifteen noise injection techniques, eight integrated baseline models pertaining to two video understanding tasks, and can be easily extend to user defined tasks, noise configurations and models. Moreover, OpenVNA provide researchers a GUI-based interface to interactively analyze the local model behavior under user specific noisy scenario. The contributions of this work can be succinctly summarized as follows,

\begin{enumerate}
    \item The OpenVNA contains one of the most comprehensive video noise injection toolkits, covering the most cases in real world applications. 
    
    \item The OpenVNA framework serves as a robust MLU benchmark, which providing unified noisy dataset construction, benchmark model reproducing and global robustness evaluation. The modular pipeline makes it very easy to integrate new models for a reliable comparison with existing baselines. 
    \item The OpenVNA framework offers a GUI-based interface, facilitating users to effortlessly apply user-defined noise to a given video and compare extracted features and model predictions, thus enabling the analysis of local model behavior and even model diagnostics.
    
\end{enumerate}

\section{Related Works}

\noindent\textbf{MultiBench.} MultiBench \cite{multibench} is a comprehensive benchmark that presents three fundamental challenges in multimodal representation learning, including generalization, complexity, and robustness. Concerning robustness, MultiBench initially outlines potential perturbations across diverse heterogeneous sources and introduces several criteria for measuring robustness. However, MultiBench primarily treats video resources as time-series and restricts the discussion to feature-level imperfections, overlooking general raw video-level perturbations. In this study, we concentrate on perturbations in video applications and further extend the previous quantitative robustness criteria to encompass all types of video noise.


\begin{table*}[ht]
\centering
\small
\begin{tabular}{lll}
\toprule[1pt]
& \textbf{Noise Type}  & \textbf{Fine-grained Noise Category} \\
\midrule[1pt]
\multirow{3}{*}{\textbf{Text}} 
& Erasing & Word Erasing Attack.\\
& Replacement & Word Replacement Attack.\\
& ASR Error       & Automatic translation error using wav2vec2-large \cite{grosman2021xlsr53-large-english}. \\ 
\midrule[1pt]
\multirow{4}{*}{\textbf{Audio}} 
& Mute and Insulation & Low-pass Filter and Volume Attenuation.\\
& Reverberation       & Hall Equalization and Room Equalization. \\ 
& Color Noise &  White, Pink, Brown, Blue, Violet and Velvet Noise.  \\ 
& Scenarios Noise & Sudden Noise and Background Noise (traffic and music, etc.)  \\ 
\midrule[1pt]
\multirow{4}{*}{\textbf{Video}} 
& Visual Occlusion   & Partial Black Draw-box and Entire Black Screen. \\ 
& Visual Blurriness  & Gaussian Blur and Average Blur. \\ 

& Noises in Digital Images & Gaussian Additive Noise. \\ 
& Visual Noise on Color Space   & Contrast, Brightness, Saturation Adjustment, Color Inversion, Channel Switching. \\ 
\bottomrule[1pt]
\end{tabular}
\caption{Types of raw data level noise supported in the OpenVNA framework. In general, eight categories of raw video level noise with more than twenty fine-grained noise is covered in the OpenVNA.}
\label{tab: noise}
\end{table*}

\begin{table*}[]
\centering
\small
\begin{tabular}{ccc}
\toprule[1pt]
& \textbf{Categories}  & \textbf{Integrated Dataset and Methods} \\
\midrule[1pt]
\multirow{2}{*}{\textbf{Datasets}} 
& Intention & MIntrec \cite{zhang2022mintrec}\\
& Sentiment & MOSI \cite{MOSI}, MOSEI \cite{MOSEI}, SIMS v2 \cite{liu2022make}\\
\midrule[1pt]
\multirow{3}{*}{\textbf{Methods}} 
& Tensor Regularization & T2FN \cite{T2FN}, TPFN \cite{TPFN}\\
& Reconstruction & TFR-Net \cite{TFR-Net}, NIAT \cite{yuan2023noise}, EMT-DFLR \cite{sun2023efficient} \\ 
& Translation &  CTFN\cite{cTFN}, MMIN \cite{MMIN}, GCNet \cite{gcnet}  \\ 
\bottomrule[1pt]
\end{tabular}
\caption{Integrated Datasets and Methods in OpenVNA.}
\label{tab: datamethod}
\end{table*}

\noindent\textbf{Robust-MSA.} The Robust-MSA \cite{Robust-MSA} is developed primarily as a demonstration platform to showcase the impact of raw video-level perturbations on MSA models. The instance level evaluation platform integrated in OpenVNA is carefully enhanced, derives partially from the Robust-MSA system. It now provides support for a broader spectrum of perturbation types, allows for custom noise injection in JSON format, and includes other notable advancements to enhance its capabilities. Furthermore, the OpenVNA framework presented in this paper strives for a more comprehensive scope by establishing a standardized evaluation process, facilitating seamless integration of newly developed models, and providing noise generation APIs to enable large-scale data generation with real-world imperfections, specifically tailored for human-centered video understanding applications.

\section{System Design}

In this section, we present the functionality of the noise injection toolkit, the global robustness evaluation benchmark, and the GUI-based interface for local behavior analysis. We commence with the noise injection toolkit, as it occupies a central position within the OpenVNA.

\subsection{Noise Injection Toolkit}

The noise injection module is designed to processes raw video input in accordance with the provided configuration. For raw data level noise, we implement a \textit{real\_noise} function to generate user-specific noise according to the list of noise items indicating the type of noise, start time, end time, and degree of noise. Based on the above function, a class named $\textit{RealNoiseConfig}$ is implemented to generate random noise configurations given alternative noise types and intensity levels. This class serves the purpose of generating noisy test data or facilitating noise-based augmentation. OpenVNA is capable of adapting to different video formats for input, and processing different video resolutions and duration, efficiently completing original video processing. For implementation, FFmpeg\footnote{https://www.ffmpeg.org/} library is utilized for video format conversion and video editing.

\noindent \textbf{Supported Noise.} Human-centered video applications naturally contains three distinct modalities: audio, visual, and textual modalities. At feature level, all three modality can be regraded as extracted feature sequence, therefore existing perturbations on time series data are considered, including random feature drop (random erasure of feature sequences with zero-padding vectors) and structural feature drop (erasing of feature sequences with zero-padding vectors in succession) \cite{yuan2023noise, multibench}. While at raw data level, audio, visual, and text are heterogeneous, different structure of noise should be considered separately. For textual modality, as the spoken words are commonly obtained from the audio modality using the Automatic Speech Recognition (ASR) technique, ASR error\footnote{ASR errors resulting from the injected audio modality noise are also taken into account.} becomes the most common noise, and supported in OpenVNA system. Besides, attacks on text including word erasing or word replacement are also supported in OpenVNA APIs. For the audio modality, four types of noise, simulating various noise sources, are considered, including mute and insulation, reverberation, color noise in laboratory environment, and additive real-world scenarios noise. For visual modality, occlusion, blurriness, noises in digital images, and noise on color space are supported. Table \ref{tab: noise} summarizes the supported raw data level noise and provides brief description for each type of noise, while detailed introduction can be found in Appendix \ref{sec: sn}.

\subsection{Global Robustness Benchmark}

OpenVNA offers researchers a unified pipeline for robust MLU model training, comprising the dataset module, method module, and evaluation module. Here, we will introduce each module individually.

\noindent \textbf{Dataset Module.} The dataset module furnishes a unified data loader interface for each supported dataset with high extensibility. Users can specific the used dataset for model reproduction using the -\-dataset command line argument. As shown in Table \ref{tab: datamethod}, OpenVNA now encompasses two specific downstream tasks, namely multimodal sentiment analysis (MSA) and multimodal intention recognition (MIR). For the MSA task, it offers support for CMU-MOSI \cite{MOSI}, CMU-MOSEI \cite{MOSEI} in English, as well as CH-SIMS v2 \cite{liu2022make} in Chinese.  As for the MIR task, it also integrates the Mintrec dataset \cite{zhang2022mintrec}. Detailed description of the above integrated databases can be found in Appendix \ref{app: id}. Additionally, it is noteworthy that the constructed noisy instances are retained within the overall databases to facilitate noise-based data augmentation.


\noindent \textbf{Method Module.} The method module offers a unified interface for model construction. Users can specific the used approach using the -\--model command line argument. Besides performing model training on original MLU datasets, OpenVNA provides optional robust training technique with generated noisy training data. By utilizing the -\-augmentation argument, the framework will additionally load the constructed noisy training instances and treat them as augmented data during the training process. Currently, as summarized in Table \ref{tab: datamethod}, OpenVNA contains eight robust MLU methods and can be roughly segmented into the tensor regularization based methods \cite{T2FN, TPFN}, the reconstruction based methods \cite{TFR-Net, sun2023efficient, yuan2023noise} and the translation based method \cite{MMIN, cTFN, gcnet}. Detailed description of all above baselines are presented in Appendix \ref{app: ib}.

\noindent \textbf{Evaluation Module.} The batch-level evaluation module of OpenVNA is devised with the aim of offering a thorough quantitative performance comparison between various methods under certain type of noise. For quantitative comparison, OpenVNA amalgamates the model performance across varying degrees of noise, offering a holistic evaluation of the robustness pertaining to the given type of noise. Specifically, OpenVNA utilizes Arbitrary Interval Robustness (AIR) as evaluation metrics, 

\begin{equation}
\gamma_{\mathtt{abs}}(f) = \int_{\sigma_{\min}}^{\sigma_{\max}} acc_\sigma(f)d\sigma, 
\end{equation}
where $[\sigma_{\min}, \sigma_{\max}]$ denotes the range of considered imperfection levels, which may vary for different types of video perturbation. This criteria geometrically evaluates the area under the accuracy-imperfection curve. In default evaluation process in OpenVNA, the integral is approximately calculated by uniformly taking the function values of 10 interior points on the interval. Besides quantitative results, performance curve are also provided for intuitive demonstration.

\subsection{Local Robustness Interface}

The instance-level evaluation module is an GUI-based web platform designed to provide a user-friendly interface for intuitive noise injection, error case studies, and even model diagnosis. It facilitates an intuitive comparison of the extracted modality features and model predictions between the original and noisy video clips. In terms of implementation, the frontend of the platform is constructed using Vue 3.0, while the backend is crafted using the Flask library in Python. Detailed installation instructions for the platform are provided on GitHub to enhance user accessibility.

\begin{table*}[ht]
\centering
\small
\begin{tabular}{c|cc|cc|cc|cc|cc|cc}
\toprule[1pt]
\multirow{2}{*}{Model} & \multicolumn{2}{c|}{R-Drop}   & \multicolumn{2}{c|}{S-Drop} & \multicolumn{2}{c|}{G-Blur} & \multicolumn{2}{c|}{Impulse} & \multicolumn{2}{c|}{Color-W} & \multicolumn{2}{c}{BG-Park} \\
& Acc-2 & F1 & Acc-2 & F1 & Acc-2 & F1 & Acc-2 & F1 & Acc-2 & F1 & Acc-2 & F1 \\
\midrule[1pt]
TPFN & 64.06 & 62.96 & 64.76 & 64.04 & 76.91 & 76.81 & 76.77 & 76.63 & 61.95 & 59.61 & 62.19 & 59.57 \\
T2FN  & 64.97 & 63.70 & 66.56 & 65.88 & 77.61 & 77.58 & 77.33 & 77.14 & 62.63 & 60.88 & 62.07 & 60.21 \\
MMIN & 63.33 & 62.23 & 65.90 & 65.45 & 76.47 & 76.53 & 76.34 & 76.42 & 60.31 & 59.90 & 59.32 & 58.87 \\
CTFN & 62.35 & 60.09 & 63.48 & 61.86 & 76.87 & 76.88 & 76.98 & 76.98 & 63.13 & 61.50 & 62.04 & 60.61 \\
GCNET & 63.84 & 63.01 & 64.78 & 62.80 & 76.44 & 76.10 & 76.35 & 76.30 & 59.10 & 58.58 & 59.76 & 58.92 \\
\midrule[0.6pt]
TPFN$^\star$ & 66.89 & 63.06 & 67.54 & 65.56 & 76.34 & 76.35 & 77.30 & 77.22 & 63.39 & 61.90 & 62.74 & 60.05 \\
T2FN$^\star$ & 65.70 & 64.21 & 66.15 & 62.11 & 76.34 & 76.35 & 76.23 & 76.24 & 63.02 & 59.26 & 62.16 & 59.29 \\
MMIN$^\star$ & 66.76 & 64.74 & 68.19 & 66.09 & 76.31 & 76.29 & 76.84 & 76.78 & 62.62 & 61.54 & 61.72 & 61.55 \\
CTFN$^\star$ & 66.54 & 65.05 & 66.73 & 66.04 & 77.14 & 77.09 & 76.82 & 76.79 & 63.32 & 61.27 & 62.94 & 61.23 \\
GCNET$^\star$ & 66.32 & 63.70 & 67.44 & 63.80 & 76.26 & 76.17 & 75.47 & 75.23 & 62.05 & 59.72 & 61.34 & 60.27 \\
TFR-Net$^\star$ & 67.47 & 65.93 & 67.55 & 66.84 & 76.06 & 76.03 & 76.36 & 76.31 & 63.07 & 61.81 & 61.80 & 61.95 \\
NIAT$^\star$ & 66.32 & 66.19 & 64.94 & 63.66 & 76.51 & 76.54 & 76.11 & 76.02 & 63.29 & 62.64 & 63.04 & 62.50 \\
EMT-DLFR$^\star$ & 67.93 & 67.22 & 68.85 & 68.37 & 77.41 & 77.45 & 76.69 & 76.80 & 63.36 & 63.34 & 63.81 & 63.85 \\
\bottomrule[1pt]
\end{tabular}
\caption{The performance of baselines on SIMS v2 dataset under random drop (denote as R-Drop), strutural drop (denote as S-Drop), Gaussian blur (denote as  G-Blur), impulse value noise (denote as Impulse), white color noise (denote as Color-W) and background noise in park (denote as BG-Park). Models marked with a $^\star$ indicate the utilization of noise-based data augmentation techniques for robust training. }
\label{tab: simsv2}
\end{table*}

\section{Framework Evaluation}

\subsection{Noise Injection Toolkit}
We present an example of injecting raw data level noise with the provided APIs below. In this example, for the visual modality, noise is randomly chosen from `Gaussian blur' and `blank', covering 80\% of the video clip with a noise intensity of 0.5. Meanwhile, for the acoustic modality, `reverberation' noise with a noise intensity of 0.3 is injected into the entire (100\%) audio waveform.

\begin{minipage}{0.94\linewidth}

\begin{lstlisting}[language=Python, caption={An example of injecting raw data level noise using OpenVNA framework.}, captionpos=b, label={lst:mmsa-fet}]
from noise_api.real_noise import real_noise, real_noise_config

cfg = real_noise_config(
    "test.mp4", 
    mode = "random_full", 
    v_noise_list = ["gblur", "blank"], 
    v_noise_num = 2,
    v_noise_ratio = 0.8,
    v_noise_intensity = 0.5,
    a_noise_list = ["reverb"],
    a_noise_num = 1,
    a_noise_ratio = 1.0,
    a_noise_intensity = 0.3,
)._asdict()

# Noise Injection with cfg.
real_noise(
    "examples/test.mp4", 
    "examples/test_out.mp4", **cfg
)

\end{lstlisting}
\end{minipage}

\begin{table}[]
\centering
\small
\begin{tabular}{ccc}
\toprule[1pt]
\textbf{Type}  & Indicator & Interval \\
\midrule[1pt]

R-Drop  & Missing Rate & [0.0, 1.0, 0.1]\\
S-Drop  & Missing Rate & [0.0, 1.0, 0.1]\\
G-Blur  & Sigma of Gaussian blur & [0, 10, 1] \\
Impulse & Strength for specific pixel & [0, 100, 10] \\
Color-W & Amplitude of the Noise & [0, 0.10, 0.01] \\
BG-Park & Amplitude of the Noise & [0.0, 1.0, 0.1]\\

\bottomrule[1pt]
\end{tabular}
\caption{Considered noise interval and brief description, where intervals are recorded in format [min, max, step].}
\label{tab: interval}
\end{table}

\subsection{Global Robustness Benchmark}

The global robustness benchmark contains both quantitative model comparison as well as the qualitative performance curve analysis. 

\noindent\textbf{Quantitative Model Comparison.} 
In Table \ref{tab: simsv2}, we recorded the AIR metrics for typical type of noise on SIMS v2 dataset. Specifically, six types of noise are considered. For feature level noise, performance under random drop (denote as \textbf{R-Drop}) and structural drop (denote as \textbf{S-Drop}) are recorded. While for raw video level noise, Gaussian blur (denote as \textbf{G-Blur}) and impulse value noise (denote as \textbf{Impulse}) are evaluated for visual, white color noise (denote as \textbf{Color-W}) and background noise in park (denote as \textbf{BG-Park}) are utilized for audio modality. The noise level intervals being considered are documented in Table \ref{tab: interval}.
Models marked with $^\star$ indicate the utilization of noise-based data augmentation techniques, where the augmented data are generated by random drop. It can be observed that model robustness can be enhanced through noise-based augmentation even for unseen type of noise. More experimental results for other datasets are provided on Github\footnote{\url{https://github.com/thuiar/OpenVNA}}.

\begin{figure}
    \centering
    \includegraphics[width=1.0\linewidth]{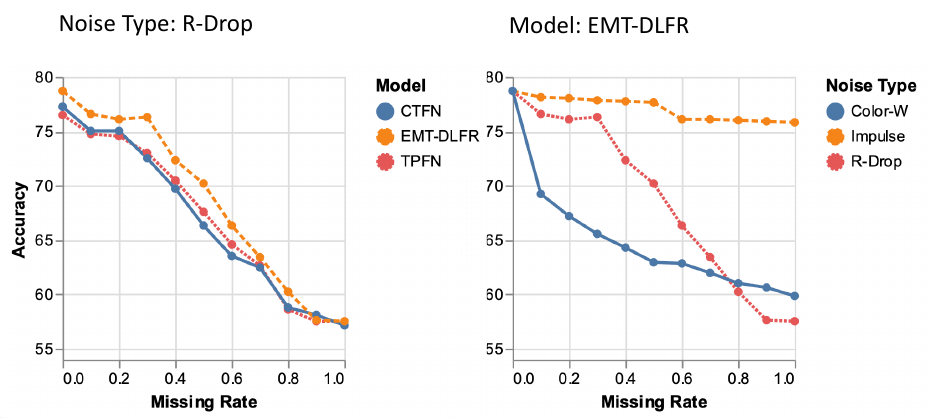}
    \caption{Fine-grained performance curve provided in OpenVNA for global robustness evaluation.}
    \label{fig: curve}
\end{figure}

\noindent\textbf{Qualitative Performance Curve.} OpenVNA also offers fine-grained performance curve comparison. As depicted in Figure \ref{fig: curve}, it allows researchers to analyze global performance from two perspectives. Firstly, comparisons can be made on the same type of noise with different models (the left sub-graph in Figure \ref{fig: curve}), providing an intuitive demonstration of how different models perform as the noise intensity increases, and aiding in model selection for various applications. Secondly, comparisons can be made on the same model with different types of noise (the right sub-graph in Figure \ref{fig: curve}), illustrating the model's sensitivity to each type of noise further enables model diagnosis and refinement.


\subsection{GUI-based Interface for Local Analysis}

\begin{figure}
    \centering
    \includegraphics[width=1.0\linewidth]{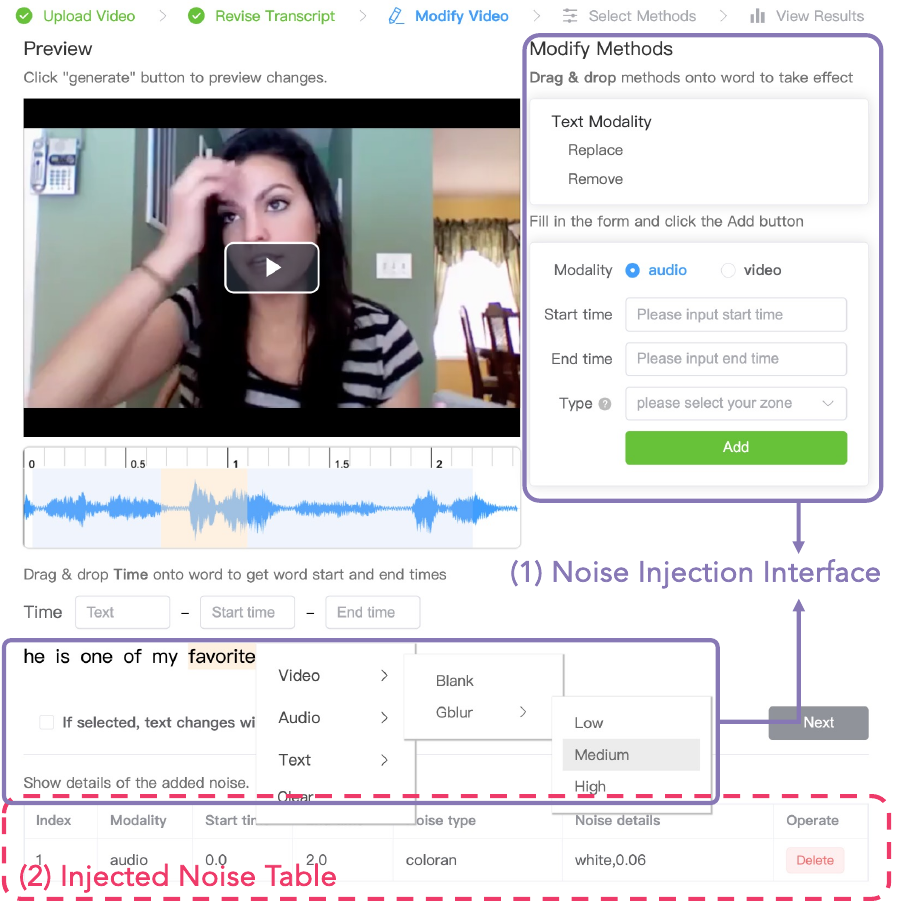}
    \caption{The GUI-based Interface of Noise Injection.}
    \label{fig: demo_1}
\end{figure}

\begin{figure}
    \centering
    \includegraphics[width=1.0\linewidth]{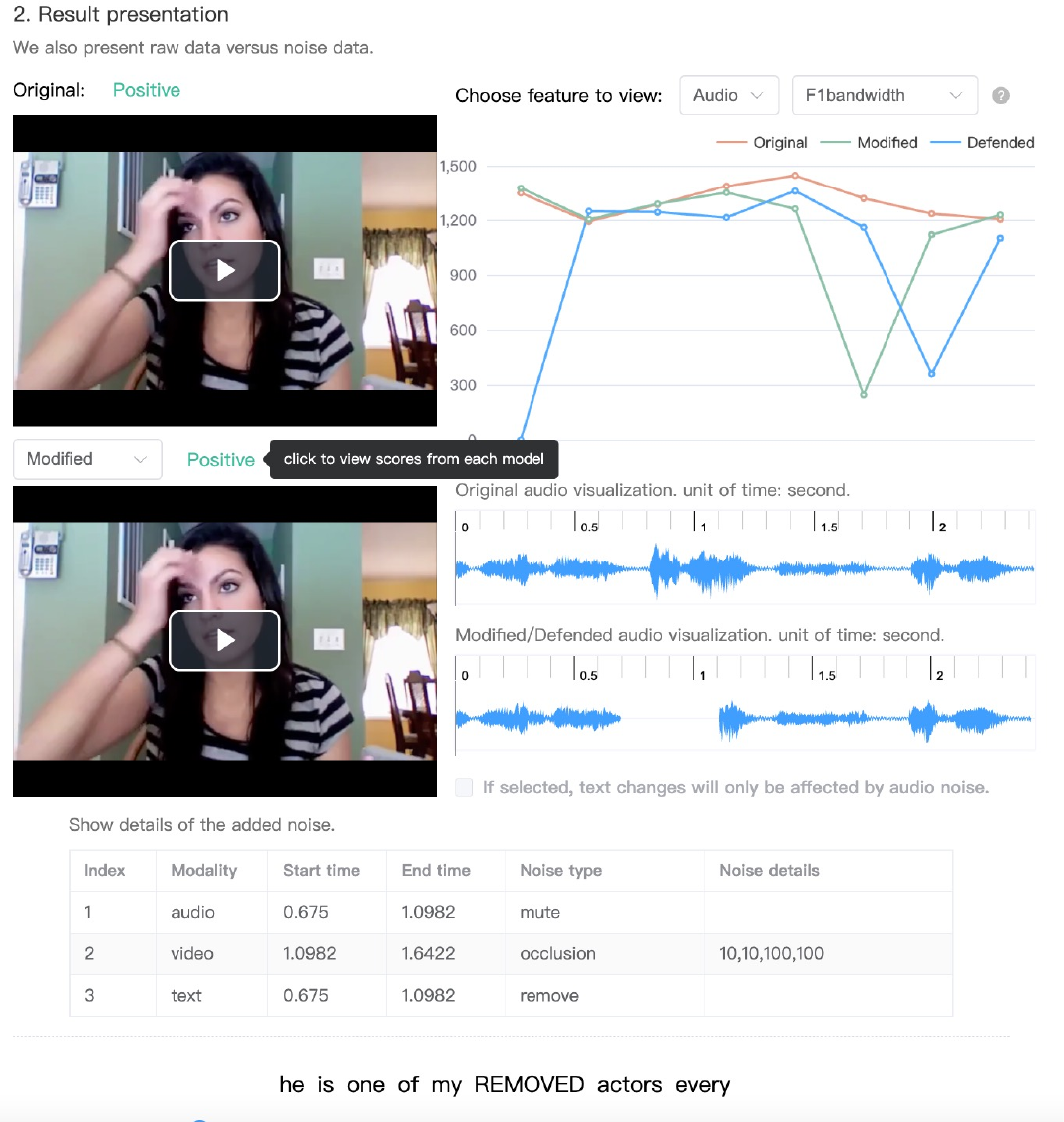}
    \caption{The GUI-based Interface of the Local model behavior analysis.}
    \label{fig: demo}
\end{figure}

The overall workflow of analyzing MLU model behavior through the provided GUI-based interface contains four main steps. Firstly, the user should upload their original video file. The Automatic Speech Recognition (ASR) technique with wav2vec-large \cite{grosman2021xlsr53-large-english} is employed to generate spoken words. Following generation process, users can manually edit and correct the ASR outputs. Based on the revised transcript, CTC segmentation \cite{ctcsegmentation} is used to find utterance alignments within the audio files. Secondly, users can perform customized noise injection either by completing noise configuration forms or by directly applying specific noise onto the video. We provide the GUI-based interface of the noise injection step in Figure \ref{fig: demo_1}. The provided interface provides the boundaries of each recognized words and thus supports injecting aligned modality noise. After editing the noise configuration, the selected noise item will be found below in the injected noise table. The noise injection is processed after clicking the generate button. User can preview the generated noisy instance before performing local analysis. The third step becomes the selection of the evaluation model and optional denoising technique. Finally, the comparison of extracted feature sequences as well as the model prediction is demonstrated for error cases analysis and causality analysis. An example of the demonstration is shown in Figure \ref{fig: demo}. 

\section{Conclusion}

In this work, we introduce OpenVNA, an open-source framework tailored for analyzing the behavior of multimodal language understanding systems under noisy conditions. This developed framework facilitates future researchers in two key ways. Firstly, OpenVNA serves as a highly extensible global robustness evaluation benchmark, integrating two video understanding tasks, four databases, and eight robust baselines. With a unified evaluation pipeline, convenient baseline reproduction is achievable, which enables a fair performance comparison. Moreover, with flexible noise injection toolkits, the provided pipeline further empowers researchers to assess designed models under analogous noise, which can be regarded as a superior simulation for real-world application scenarios. Secondly, OpenVNA provides a GUI-based interface for local model behavior analysis. Through detailed comparisons of extracted feature sequences and model predictions, ad-hoc model behavior explanations can be formulated, facilitating error case analysis and model diagnostics. The authors firmly believe that OpenVNA will make a significant contribution to the advancement of robust multimodal applications and foster future research endeavors.



\section*{Acknowledgements}

This work is founded by National Natural Science Foundation of China (Grant No. 62173195), National Science and Technology Major Project towards the new generation of broadband wireless mobile communication networks of Jiangxi Province (03 and 5G Major Project of Jiangxi Province, Grant No. 20232ABC03A02), and Natural Science Foundation of Hebei Province, China (Grant No. F2022208006).



\begin{thebibliography}{38}
\expandafter\ifx\csname natexlab\endcsname\relax\def\natexlab#1{#1}\fi

\bibitem[{Baltru{\v{s}}aitis et~al.(2018)Baltru{\v{s}}aitis, Ahuja, and Morency}]{baltruvsaitis2018multimodal}
Tadas Baltru{\v{s}}aitis, Chaitanya Ahuja, and Louis-Philippe Morency. 2018.
\newblock Multimodal machine learning: A survey and taxonomy.
\newblock \emph{IEEE transactions on pattern analysis and machine intelligence}, 41(2):423--443.

\bibitem[{Baltru{\v{s}}aitis et~al.(2016)Baltru{\v{s}}aitis, Robinson, and Morency}]{baltruvsaitis2016openface}
Tadas Baltru{\v{s}}aitis, Peter Robinson, and Louis-Philippe Morency. 2016.
\newblock Openface: an open source facial behavior analysis toolkit.
\newblock In \emph{2016 IEEE winter conference on applications of computer vision (WACV)}, pages 1--10. IEEE.

\bibitem[{Eyben et~al.(2015)Eyben, Scherer, Schuller, Sundberg, Andr{\'e}, Busso, Devillers, Epps, Laukka, Narayanan et~al.}]{eyben2015geneva}
Florian Eyben, Klaus~R Scherer, Bj{\"o}rn~W Schuller, Johan Sundberg, Elisabeth Andr{\'e}, Carlos Busso, Laurence~Y Devillers, Julien Epps, Petri Laukka, Shrikanth~S Narayanan, et~al. 2015.
\newblock The geneva minimalistic acoustic parameter set (gemaps) for voice research and affective computing.
\newblock \emph{IEEE transactions on affective computing}, 7(2):190--202.

\bibitem[{Grosman(2021)}]{grosman2021xlsr53-large-english}
Jonatas Grosman. 2021.
\newblock Fine-tuned {XLSR}-53 large model for speech recognition in {E}nglish.
\newblock \url{https://huggingface.co/jonatasgrosman/wav2vec2-large-xlsr-53-english}.

\bibitem[{Hazarika et~al.(2022)Hazarika, Li, Cheng, Zhao, Zimmermann, and Poria}]{hazarika2022analyzing}
Devamanyu Hazarika, Yingting Li, Bo~Cheng, Shuai Zhao, Roger Zimmermann, and Soujanya Poria. 2022.
\newblock Analyzing modality robustness in multimodal sentiment analysis.
\newblock \emph{arXiv preprint arXiv:2205.15465}.

\bibitem[{Kenton and Toutanova(2019)}]{bert}
Jacob Devlin Ming-Wei~Chang Kenton and Lee~Kristina Toutanova. 2019.
\newblock Bert: Pre-training of deep bidirectional transformers for language understanding.
\newblock In \emph{Proceedings of naacL-HLT}, volume~1, page~2.

\bibitem[{K{\"u}rzinger et~al.(2020)K{\"u}rzinger, Winkelbauer, Li, Watzel, and Rigoll}]{ctcsegmentation}
Ludwig K{\"u}rzinger, Dominik Winkelbauer, Lujun Li, Tobias Watzel, and Gerhard Rigoll. 2020.
\newblock Ctc-segmentation of large corpora for german end-to-end speech recognition.
\newblock In \emph{Speech and Computer}, pages 267--278, Cham. Springer International Publishing.

\bibitem[{Li et~al.(2020)Li, Li, Duan, Zheng, and Zhao}]{TPFN}
Binghua Li, Chao Li, Feng Duan, Ning Zheng, and Qibin Zhao. 2020.
\newblock Tpfn: Applying outer product along time to multimodal sentiment analysis fusion on incomplete data.
\newblock In \emph{European Conference on Computer Vision}, pages 431--447. Springer.

\bibitem[{Li et~al.(2023)Li, He, Wang, Li, Wang, Luo, Wang, Wang, and Qiao}]{li2023videochat}
KunChang Li, Yinan He, Yi~Wang, Yizhuo Li, Wenhai Wang, Ping Luo, Yali Wang, Limin Wang, and Yu~Qiao. 2023.
\newblock Videochat: Chat-centric video understanding.
\newblock \emph{arXiv preprint arXiv:2305.06355}.

\bibitem[{Lian et~al.(2023{\natexlab{a}})Lian, Chen, Sun, Liu, and Tao}]{gcnet}
Zheng Lian, Lan Chen, Licai Sun, Bin Liu, and Jianhua Tao. 2023{\natexlab{a}}.
\newblock Gcnet: graph completion network for incomplete multimodal learning in conversation.
\newblock \emph{IEEE Transactions on Pattern Analysis and Machine Intelligence}.

\bibitem[{Liang et~al.(2021)Liang, Lyu, Fan, Wu, Cheng, Wu, Chen, Wu, Lee, Zhu et~al.}]{multibench}
P~Liang, Y~Lyu, X~Fan, Z~Wu, Y~Cheng, J~Wu, LY~Chen, P~Wu, MA~Lee, Y~Zhu, et~al. 2021.
\newblock Multibench: Multiscale benchmarks for multimodal representation learning.
\newblock In \emph{In Proceedings of the Neural Information Processing Systems Conference (Neurips)}.

\bibitem[{Liang et~al.(2019)Liang, Liu, Tsai, Zhao, Salakhutdinov, and Morency}]{T2FN}
Paul~Pu Liang, Zhun Liu, Yao-Hung~Hubert Tsai, Qibin Zhao, Ruslan Salakhutdinov, and Louis-Philippe Morency. 2019.
\newblock Learning representations from imperfect time series data via tensor rank regularization.
\newblock In \emph{Proceedings of the 57th Annual Meeting of the Association for Computational Linguistics}, pages 1569--1576.

\bibitem[{Liang et~al.(2022)Liang, Zadeh, and Morency}]{liang2022foundations}
Paul~Pu Liang, Amir Zadeh, and Louis-Philippe Morency. 2022.
\newblock Foundations and recent trends in multimodal machine learning: Principles, challenges, and open questions.
\newblock \emph{arXiv preprint arXiv:2209.03430}.

\bibitem[{Liu et~al.(2022)Liu, Yuan, Mao, Liang, Yang, Qiu, Cheng, Li, Xu, and Gao}]{liu2022make}
Yihe Liu, Ziqi Yuan, Huisheng Mao, Zhiyun Liang, Wanqiuyue Yang, Yuanzhe Qiu, Tie Cheng, Xiaoteng Li, Hua Xu, and Kai Gao. 2022.
\newblock Make acoustic and visual cues matter: Ch-sims v2. 0 dataset and av-mixup consistent module.
\newblock In \emph{Proceedings of the 2022 International Conference on Multimodal Interaction}, pages 247--258.

\bibitem[{Ma et~al.(2022)Ma, Ren, Zhao, Testuggine, and Peng}]{ma2022multimodal}
Mengmeng Ma, Jian Ren, Long Zhao, Davide Testuggine, and Xi~Peng. 2022.
\newblock Are multimodal transformers robust to missing modality?
\newblock In \emph{Proceedings of the IEEE/CVF Conference on Computer Vision and Pattern Recognition}, pages 18177--18186.

\bibitem[{Maaz et~al.(2023)Maaz, Rasheed, Khan, and Khan}]{video_chatgpt}
Muhammad Maaz, Hanoona Rasheed, Salman Khan, and Fahad~Shahbaz Khan. 2023.
\newblock Video-chatgpt: Towards detailed video understanding via large vision and language models.
\newblock \emph{arXiv preprint arXiv:2306.05424}.

\bibitem[{Mao et~al.(2022{\natexlab{a}})Mao, Yuan, Xu, Yu, Liu, and Gao}]{mao2022m}
Huisheng Mao, Ziqi Yuan, Hua Xu, Wenmeng Yu, Yihe Liu, and Kai Gao. 2022{\natexlab{a}}.
\newblock M-sena: An integrated platform for multimodal sentiment analysis.
\newblock In \emph{Proceedings of the 60th Annual Meeting of the Association for Computational Linguistics: System Demonstrations}, pages 204--213.

\bibitem[{Mao et~al.(2022{\natexlab{b}})Mao, Zhang, Xu, Yuan, and Liu}]{Robust-MSA}
Huisheng Mao, Baozheng Zhang, Hua Xu, Ziqi Yuan, and Yihe Liu. 2022{\natexlab{b}}.
\newblock Robust-msa: Understanding the impact of modality noise on multimodal sentiment analysis.
\newblock \emph{arXiv preprint arXiv:2211.13484}.

\bibitem[{Soleymani et~al.(2017)Soleymani, Garcia, Jou, Schuller, Chang, and Pantic}]{soleymani2017survey}
Mohammad Soleymani, David Garcia, Brendan Jou, Bj{\"o}rn Schuller, Shih-Fu Chang, and Maja Pantic. 2017.
\newblock A survey of multimodal sentiment analysis.
\newblock \emph{Image and Vision Computing}, 65:3--14.

\bibitem[{Sun et~al.(2023)Sun, Lian, Liu, and Tao}]{sun2023efficient}
Licai Sun, Zheng Lian, Bin Liu, and Jianhua Tao. 2023.
\newblock Efficient multimodal transformer with dual-level feature restoration for robust multimodal sentiment analysis.
\newblock \emph{IEEE Transactions on Affective Computing}.

\bibitem[{Tang et~al.(2021)Tang, Li, Jin, Cichocki, Zhao, and Kong}]{cTFN}
Jiajia Tang, Kang Li, Xuanyu Jin, Andrzej Cichocki, Qibin Zhao, and Wanzeng Kong. 2021.
\newblock Ctfn: Hierarchical learning for multimodal sentiment analysis using coupled-translation fusion network.
\newblock In \emph{Proceedings of the 59th Annual Meeting of the Association for Computational Linguistics and the 11th International Joint Conference on Natural Language Processing (Volume 1: Long Papers)}, pages 5301--5311.

\bibitem[{Yuan et~al.(2021)Yuan, Li, Xu, and Yu}]{TFR-Net}
Ziqi Yuan, Wei Li, Hua Xu, and Wenmeng Yu. 2021.
\newblock Transformer-based feature reconstruction network for robust multimodal sentiment analysis.
\newblock In \emph{Proceedings of the 29th ACM International Conference on Multimedia}, pages 4400--4407.

\bibitem[{Yuan et~al.(2023)Yuan, Liu, Xu, and Gao}]{yuan2023noise}
Ziqi Yuan, Yihe Liu, Hua Xu, and Kai Gao. 2023.
\newblock Noise imitation based adversarial training for robust multimodal sentiment analysis.
\newblock \emph{IEEE Transactions on Multimedia}.

\bibitem[{Zadeh et~al.(2016)Zadeh, Zellers, Pincus, and Morency}]{MOSI}
Amir Zadeh, Rowan Zellers, Eli Pincus, and Louis-Philippe Morency. 2016.
\newblock Multimodal sentiment intensity analysis in videos: Facial gestures and verbal messages.
\newblock \emph{IEEE Intelligent Systems}, 31(6):82--88.

\bibitem[{Zadeh et~al.(2018)Zadeh, Liang, Poria, Cambria, and Morency}]{MOSEI}
AmirAli~Bagher Zadeh, Paul~Pu Liang, Soujanya Poria, Erik Cambria, and Louis-Philippe Morency. 2018.
\newblock Multimodal language analysis in the wild: Cmu-mosei dataset and interpretable dynamic fusion graph.
\newblock In \emph{Proceedings of the 56th Annual Meeting of the Association for Computational Linguistics (Volume 1: Long Papers)}, pages 2236--2246.

\bibitem[{Zhang et~al.(2023)Zhang, Li, and Bing}]{zhang2023video}
Hang Zhang, Xin Li, and Lidong Bing. 2023.
\newblock Video-llama: An instruction-tuned audio-visual language model for video understanding.
\newblock \emph{arXiv preprint arXiv:2306.02858}.

\bibitem[{Zhang et~al.(2022)Zhang, Xu, Wang, Zhou, Zhao, and Teng}]{zhang2022mintrec}
Hanlei Zhang, Hua Xu, Xin Wang, Qianrui Zhou, Shaojie Zhao, and Jiayan Teng. 2022.
\newblock Mintrec: A new dataset for multimodal intent recognition.
\newblock In \emph{Proceedings of the 30th ACM International Conference on Multimedia}, pages 1688--1697.

\bibitem[{Zhao et~al.(2021)Zhao, Li, and Jin}]{MMIN}
Jinming Zhao, Ruichen Li, and Qin Jin. 2021.
\newblock Missing modality imagination network for emotion recognition with uncertain missing modalities.
\newblock In \emph{Proceedings of the 59th Annual Meeting of the Association for Computational Linguistics and the 11th International Joint Conference on Natural Language Processing (Volume 1: Long Papers)}, pages 2608--2618.

\end{thebibliography}

\bibliographystyle{acl_natbib}

\appendix

\appendix

\begin{table*}[]
\centering
\small
\begin{tabular}{c|cc|cc|cc|cc|cc|cc}
\toprule[1pt]
\multirow{2}{*}{Model} & \multicolumn{2}{c|}{R-Drop}   & \multicolumn{2}{c|}{S-Drop} & \multicolumn{2}{c|}{G-Blur} & \multicolumn{2}{c|}{Impulse} & \multicolumn{2}{c|}{Color-W} & \multicolumn{2}{c}{BG-Park} \\
& Acc-2 & F1 & Acc-2 & F1 & Acc-2 & F1 & Acc-2 & F1 & Acc-2 & F1 & Acc-2 & F1 \\
\midrule[1pt]
TPFN & 64.26 & 56.22 & 65.69 & 61.48 & 78.81 & 78.92 & 79.85 & 79.90 & 62.17 & 61.82 & 63.12 & 62.94 \\
T2FN  & 64.07 & 59.19 & 63.87 & 59.92 & 79.14 & 79.19 & 79.43 & 79.40 & 64.72 & 64.64 & 65.23 & 65.24 \\
MMIN & 64.89 & 56.64 & 66.66 & 63.05 & 79.42 & 79.49 & 80.58 & 80.53 & 62.94 & 62.65 & 63.57 & 63.41 \\
CTFN & 65.52 & 57.76 & 67.61 & 63.92 & 80.02 & 80.05 & 79.44 & 79.51 & 66.34 & 66.19 & 66.39 & 66.19 \\
GCNET & 64.23 & 60.40 & 64.12 & 56.11 & 78.86 & 78.90 & 78.33 & 78.26 & 64.18 & 64.29 & 65.11 & 65.25 \\
\midrule[0.6pt]
TPFN$^\star$ & 63.61 & 61.18 & 66.70 & 65.72 & 79.83 & 79.88 & 78.98 & 78.95 & 62.62 & 62.46 & 63.17 & 62.90 \\
T2FN$^\star$ & 63.69 & 62.46 & 64.22 & 63.74 & 78.96 & 79.02 & 79.32 & 79.32 & 62.48 & 62.64 & 64.75 & 64.63 \\
MMIN$^\star$ & 65.53 & 64.39 & 67.05 & 65.31 & 80.56 & 80.49 & 81.05 & 80.99 & 65.57 & 65.00 & 67.62 & 66.99 \\
CTFN$^\star$ & 65.60 & 63.63 & 64.93 & 64.53 & 78.43 & 78.54 & 79.21 & 79.28 & 64.89 & 64.92 & 66.01 & 65.85 \\
GCNET$^\star$ & 62.98 & 61.51 & 64.76 & 63.75 & 76.46 & 76.59 & 78.48 & 78.38 & 65.58 & 65.59 & 65.07 & 64.89 \\
TFR-Net$^\star$& 67.39 & 66.48 & 66.60 & 64.90 & 81.88 & 81.92 & 82.20 & 82.30 & 64.98 & 64.93 & 67.47 & 67.35 \\
NIAT$^\star$ & 67.92 & 67.19 & 70.65 & 70.23 & 83.66 & 83.67 & 84.27 & 84.15 & 66.29 & 65.98 & 66.87 & 67.01 \\
EMT-DLFR$^\star$ & 68.67 & 67.51 & 71.00 & 70.79 & 84.16 & 84.17 & 84.79 & 84.73 & 66.48 & 66.55 & 65.10 & 64.85 \\
\bottomrule[1pt]
\end{tabular}
\caption{The performance of selected baselines on MOSI dataset. $^\star$ indicates that data augmentation is applied, and the augmentation type is consistent with the validation type.}
\label{tab: mosi}
\end{table*}

\section{Details of Supported Noise}
\label{sec: sn}
Detailed description of raw data level audio noise is provided as follows,

\noindent \textbf{Emulation of Mute and Insulation.} Due to the occurrence of insulation or translation errors, some voice components might be lost in the recorded audio wave form. A low-pass filter is employed to replicate the insulation effect, as high-frequency components are more susceptible to insulation. Additionally, a mute mode is incorporated to simulate the translation error and severe volume attenuation.

\noindent \textbf{Emulation of Reverberation.}
Reverberation is a common speech phenomenon that arises within enclosed spaces, resulting from the superposition of direct and reflected sounds. This study involves simulating two archetypal forms of reverberation, namely hall and room equalization, by employing finite impulse response filters with pre-established reverberation hyperparameters.

\noindent \textbf{Color Noise in Laboratory Environment.}
Color noise is a series of meticulously crafted laboratory-generated noise, stemming from the domain of psychological acoustics. It provides researchers with an ideal and controllable emulation of various environmental noises. Within this study, we have amalgamated six common types of color noise - white, pink, brown, blue, violet, and velvet noise - and blended them with the original speech to assess the overall robustness of the model. For an elaborate elucidation and depiction of each variant of color noise, comprehensive information can be found on the demo website, and Github.

\noindent \textbf{Real-world Scenarios Noise.}
In addition to the ideal simulations, this study also presents nine distinct real-world recordings of acoustic noise from various environments, such as noise captured in parks, restaurants, and others. A comprehensive description and instances can be accessed on the public website. The real-world noise scenarios offered are properly combined with the original speech to effectively demonstrate the potential impacts on downstream video understanding tasks when exposed to such types of noise.

Detailed description for raw data level visual noise is provided as follows,

\noindent \textbf{Visual Occlusion.} Video clips in real-world applications may encounter occlusion in certain parts of the video region. The OpenVNA framework introduces occlusion by overlaying a black draw-box to cover the designated region.

\noindent \textbf{Visual Blurriness.} The most common types of blur include Gaussian blur, box blur, variable blur and radial blur. The developed OpenVNA implements Gaussian blur, which emulates a "frosted glass" effect, and the box blur,   which imitates the Bokeh effect of a single-lens reflex camera.

\noindent \textbf{Noises in Digital Images.} This type of noise is naturally prevalent in digital images during image acquisition, coding, transmission, and processing stages. The OpenVNA framework incorporates Gaussian additive noise that normalizes the histogram concerning the gray values.

\noindent \textbf{Visual Noise on Color Space.} To introduce noise in the color space, the OpenVNA framework offers strategies for contrast, brightness, saturation, and gamma adjustments, effectively simulating diverse illumination environments using color filters. Besides, color inversion and channel switching (e.g., from 'RGB' to 'BGR') are also integrated.

\section{Details of Integrated Databases}
\label{app: id}
The framework offers support for three distinguished benchmark datasets, namely MOSI, MOSEI, and CH-SIMS v2, meticulously curated for Multimodal Sentiment Analysis (MSA) endeavors. Furthermore, it encompasses the MIntRec dataset, designed to cater to the domain of Multimodal Information Retrieval (MIR) tasks.

\noindent \textbf{CMU-MOSI} \cite{MOSI} is a widely used MSA dataset containing 2199 video clips from 93 YouTube movie review videos. Labels range from -3 (strongly negative) to 3 (strongly positive). 

\noindent \textbf{CMU-MOSEI} \cite{MOSEI} is an extended version of the MOSI dataset, designed to include a larger number of utterances, a wider range of samples, speakers, and topics. It consists of 23,453 annotated video segments extracted from 5,000 videos. The dataset includes utterances from 1,000 distinct speakers and covers 250 different topics.

\noindent \textbf{CH-SIMS v2} \cite{liu2022make} is a popular Chinese MSA benchmark dataset. It has doubled the size of the original CH-SIMS dataset, making it more comprehensive and diverse. Notably, this dataset has been verified to demonstrate the significance of nonverbal behaviors in predicting emotions.

\noindent \textbf{MIntRec} \cite{zhang2022mintrec} formulates intent classification based on data collected from  the TV series  Supermarkets.   The dataset consists of 2,224 high-quality samples with text, video, and audio patterns, and includes 20 intent categories.

\section{Feature Extraction}
For all experiments in this paper, the MMSA-FET toolkit \cite{mao2022m} is employed to extract unaligned features. For visual modality, we extract 35 dimensions of Action Units (AUs) as described in OpenFace \cite{baltruvsaitis2016openface} and 136 dimensions of 68 facial landmarks, at a sample rate of 10 frames per second.  For audio modality, we use the eGeMAPSv02 feature set \cite{eyben2015geneva}, which is of 25 dimensions.  For the text modality, we use BERT \cite{bert} which consists of 768 dimensions.

\section{Details of Integrated Baselines}
\label{app: ib}

\noindent \textbf{Tensor regularization based methods}: T2FN \cite{T2FN} uses tensor rank minimization to regularize the high rank caused by partial missing modalities. TPFN \cite{TPFN} takes high-order statistics over both modalities and temporal dynamics into account, and calculate outer products along time-steps. 

\noindent \textbf{Reconstruction based methods}: {TFR-Net} \cite{TFR-Net} exploits intra-modal and inter-modal attention-based extractors to learn robust representations for each element in modality sequences and then use a reconstruction module to generate the missing modality features. EMT-DLFR \cite{sun2023efficient} improve former low-level feature reconstruction with high-level feature attraction to achieve robust performance. NIAT \cite{yuan2023noise} integrates noise-aware adversarial training and utterance-level semantics reconstruction to narrow the representation gap between original and noisy data pairs.

\noindent \textbf{Translation based methods}: cTFN \cite{cTFN} models bi-directional cross-modality inter-correlation in parallel via couple learning, and establishes a hierarchical architecture to exploit multiple bi-directional translations. MMIN \cite{MMIN} learns robust joint multimodal representations via the Cascade Residual Auto-encoder and Cycle Consistency Learning. GCNET \citet{gcnet}  leverages graph neural networks to capture temporal and speaker information in conversations, aiming to learn discriminative representations from modality-incomplete conversational data.

\section{Supplementary Experiments}
\label{sec: se}
To gain a deeper understanding of the model's performance across various noise conditions, we have carefully chosen six distinct types of noise.  For feature level noise, performance under random drop (denote as \textbf{R-Drop}) and structural drop (denote as \textbf{S-Drop}) are recorded. While for raw video level noise, Gaussian blur (denote as \textbf{G-Blur}) and impulse value noise (denote as \textbf{Impulse}) are evaluated for visual, white color noise (denote as \textbf{Color-W}) and background noise in park (denote as \textbf{BG-Park}) are utilized for audio. By introducing these noise types, our objective is to evaluate how the model behaves and performs in different noisy environments. This analysis will enable us to assess the model's robustness and adaptability to various noise sources, potentially identifying areas that require improvement. Table \ref{tab: simsv2} presents the results on the SIMS v2 dataset, while Table \ref{tab: mosi} showcases the results on the MOSI dataset.

\end{document}